\documentclass[10pt]{article}

\begin{document}

\title{Levi-Civita spacetimes in multidimensional theories}
\author{J. Ponce de Leon\thanks{E-Mail:
jpdel@ltp.upr.clu.edu, jpdel1@hotmail.com}  \\Laboratory of Theoretical Physics, 
Department of Physics\\ 
University of Puerto Rico, P.O. Box 23343,  
San Juan,\\ PR 00931, USA}
\date{April 2009}

\maketitle
\begin{abstract}
We obtain the most general static cylindrically symmetric vacuum solutions of the Einstein field equations in $(4 + N)$ dimensions. Under the assumption of separation of variables, we construct a family of Levi-Civita-Kasner  vacuum  solutions in $(4 + N)$.  We discuss the dimensional reduction of the static solutions. Depending on the reduction procedure, they can be interpreted either as a scalar-vacuum generalization of Levi-Civita spacetimes, or as the effective $4D$ vacuum spacetime outside of an idealized string in braneworld theory.

\end{abstract}

\medskip



\newpage

{\bf Introduction:}  Levi-Civita spacetimes are static, cylindrically  symmetric vacuum solutions of the Einstein field equations. 
They can be written as (see, e.g. \cite{Bicak})

\begin{equation}
\label{Levi-Civita spacetime}
ds^2 =  r^{2 m}dt^2 - r^{2 m(m - 1)}\left(dr^2 + dz^2\right) - {{\cal{C}}}^2 \; r^{2(1 - m)}\; d\phi^2, 
\end{equation}
where $-\infty < t < \infty $ is the time, $0 \leq r < \infty$ is a cylindrical radial coordinate, $- \infty < z < \infty $ is the longitudinal coordinate, $\phi \in [0, 2\pi]$ is the angular one, $m$ and ${\cal{C}}$ are constant parameters related to the
local curvature and the conicity of the spacetime,  respectively. These spacetimes have been widely discussed in the literature from different standpoints, ranging from quantum singularities \cite{Konkowski1} to wormholes \cite{Bronnikovwormholes}. Although there has been  some controversy regarding the physical interpretation and range of $m$ and ${\cal{C}}$,  the Levi-Civita solution  (\ref{Levi-Civita spacetime}) is usually interpreted as the spacetime generated by infinite line masses and             
idealized cosmic strings \cite{Bicak}, \cite{Gautreau}-\cite{Herrera3}.  

Nowadays, there are a number of theories suggesting that the universe may have more than  four dimensions. In the case of starlike structures, described in the framework of spherical symmetry, 
higher dimensional extensions of the Schwarzschild metric have been used to calculate possible observational  effects of extra dimensions, which include the classical tests of relativity, as well as the geodesic precession of a gyroscope and possible departures from the equivalence principle \cite{WessonBook}-\cite{HongyaOverduin}. Such calculations are important in view of the Satellite Test of the Equivalence Principle (STEP), which  will advance experimental
limits on violations of Einstein's equivalence principle \cite{STEP}.

In addition to starlike structures, the universe may contain matter configurations  that are infinitely extended along a certain direction, like cosmic strings. Idealized cosmic strings are cylindrically symmetric. Therefore,  in the context of higher dimensional theories it seems to be  natural and necessary  to generalize  the Levi-Civita spacetimes (\ref{Levi-Civita spacetime}) to more than four dimensions.  

In this paper we accomplish this goal.  First, we obtain the most general static cylindrically symmetric vacuum solutions of the Einstein field equations in $(4 + N)$ dimensions. Second, under the assumption of separation of variables, we construct a time-dependent family of vacuum  solutions in $(4 + N)$ that generalize  (\ref{Levi-Civita spacetime}).  Third, we discuss the question of how these multidimensional solutions reduce to $4D$. 

\medskip 

{\bf The static solution:} Let us consider static, cylindrically symmetric spacetimes with the general metric in the form

\begin{equation}
\label{static metric}
dS^2 =   e^{2\gamma(u)} dt^2 - e^{2\alpha(u)} du^2 - e^{2\xi(u)} dz^2 - e^{2\beta(u)} d\phi^2 - \sum_{i = 1}^{N} e^{2 \sigma_{(i)}(u)}d l_{(i)}^2,
\end{equation}
where $u$ is an arbitrary admissible cylindrical radial coordinate,  and $l_{(i)}$ $(i = 1, 2, \cdots , N)$ represents   the coordinate  along the $i$-th extra dimension. The Einstein field equations in vacuum $R_{AB} = 0$, with $A \neq 1$ and $B \neq 1$, yield

\begin{eqnarray}
\label{field equations}
\gamma'' + \gamma' (\gamma' - \alpha' + \beta' + \xi' + \sigma') = 0, \nonumber \\
\xi'' + \xi' (\gamma' - \alpha' + \beta' + \xi' + \sigma') = 0, \nonumber \\
\beta'' + \beta' (\gamma' - \alpha' + \beta' + \xi' + \sigma') = 0, \nonumber \\
\sigma_{(i)}'' + \sigma_{(i)}' (\gamma' - \alpha' + \beta' + \xi' + \sigma') = 0, 
\end{eqnarray}
where the prime denotes $d/du$ and $\sigma \equiv \sum_{i = 1}^{N} \sigma_{(i)}$. We now choose $u$ as a harmonic radial coordinate, which is defined by the condition \cite{BronnikovHarmonic}
\begin{equation}
\label{harmonic radial coordinate}
\alpha = \beta + \gamma + \xi + \sigma.
\end{equation}
With this choice the field equations lead to
\begin{equation}
\label{static solution}
\gamma(u) = a u + a_{0}, \;\;\;\beta(u) = b u + b_{0}, \;\;\;\xi(u) = c u + c_{0}, \;\;\; \sigma_{(i)}(u) = s_{(i)} u + s_{(i)0}, 
\end{equation}
where $(a, a_{0}, b, b_{0},  c, c_{0},  s_{(i)}, s_{(i)0})$ are constants of integration. From $R_{11} = 0$ we find that these constants must satisfy the condition
\begin{equation}
\label{compatibility condition}
\left(a + b + c + s\right)^2 - a^2 - b^2 - c^2 - {\bar{s}}^2 = 0, 
\end{equation}
with
\begin{equation}
\label{definition of s and sbar}
s \equiv \sum_{i = 1}^{N} s_{(i)}, \;\;\;\; {\bar{s}}^2 \equiv \sum_{i = 1}^{N}s_{(i)}^2.
\end{equation}

Without loss of generality, we can set $a_{0} = b_{0} = c_{0} = s_{(i)0} = 0$ by changing the scales along $t$, $z$ and the $l_{(i)}$ axes and choosing the zero point of the $u$ coordinate. Thus, the metric takes the form

\begin{equation}
\label{static metric, solution}
dS^2 =  e^{2 a u} dt^2 - e^{2 (a + b + c + s) u} \; du^2  - e^{2 c u} dz^2 - e^{2 b u} d\phi^2 - \sum_{i = 1}^{N} e^{2 s_{(i)} u} \; d l_{(i)}^2.
\end{equation}

With the transformation of coordinates\footnote{From (\ref{compatibility condition}) and (\ref{definition of s and sbar}) it follows that  $(a + b + c + s) = 0$ leads to $a = b = c = s_{(i)} = 0$. Also, $(a + b + s) = 0$ implies $a = b = s_{(i)} = 0$. In both cases the spacetime is Riemann-flat. In what follows we assume $(a + b + c + s) \neq 0$ and $(a + b + s) \neq 0$.}
\begin{equation}
\label{transformation of coordinates}
e^{(a + b + s) u} = (a + b + s)\; r
\end{equation}
and the notation
\begin{eqnarray}
\label{notation}
m &=& \frac{a}{a + b + s}, \;\;\;{\cal{C}} = \left(a + b + s\right)^{b/(a + b + c + s)}, \nonumber \\
w &=& \frac{s}{a + b + s}, \;\;\; w_{(i)} = \frac{s_{(i)}}{a + b + s}, \;\;\; {\bar{w}}^2 = \sum_{i = 1}^{N} w_{(i)}^2,\nonumber \\
\left(\frac{\tilde{t}}{t}\right)^{1/a} &=& \left(\frac{\tilde{r}}{r}\right)^{1/c} = \left(\frac{\tilde{z}}{z}\right)^{1/c} = \left(\frac{{\tilde{l}}_{(i)}}{l_{(i)}}\right)^{1/s_{(i)}} = \left(a + b + s\right)^{1/(a + b + c + s)},
\end{eqnarray}
the line element (\ref{static metric, solution}) becomes 

\begin{equation}
\label{Levi-Civita in 4 + N}
dS^2 =  {\tilde{r}}^{2 m} d{\tilde{t}}^2 - {\tilde{r}}^{\left[2 (m + w)(m - 1) + w^2 + {\bar{w}}^2\right]}\left( d{\tilde{r}}^2 + d{\tilde{z}}^2\right) - {\cal{C}}^2 {\tilde{r}}^{2(1 - m - w)} d\phi^2 - \sum_{i = 1}^{N} {\tilde{r}}^{2 w_{(i)}} \; d {\tilde{l}}_{(i)}^2.
\end{equation}
We note that ${\cal{C}}$ cannot
be removed by a coordinate transformation. When all the coefficients  $w_{(i)}$ $(s_{(i)})$ are zero it reduces to the Levi-Civita solution (\ref{Levi-Civita spacetime}) with $N$ flat, and therefore innocuous, extra dimensions. Thus, (\ref{Levi-Civita in 4 + N}) generalizes  Levi-Civita's spacetimes to $(4 + N)$ dimensions. 

\medskip 

{\bf The time-dependent solution:} We now proceed to solve the field equations $R_{A B} = 0$ for the case where the metric functions are separable in $t$ and $u$, i.e., when they can be written as a product of two functions; one of $t$ and the other of $u$.   By an appropriate choice of the time coordinate the function of $t$ in  $g_{tt}$ may  be turned to one. Therefore, we consider the metric  

\begin{equation}
\label{separable metric}
dS^2 =   e^{2\gamma(u)} dt^2 - A^{2}(t) e^{2\alpha(u)} du^2 - B^2(t) e^{2\xi(u)} dz^2 - C^2(t) e^{2\beta(u)} d\phi^2                             - \sum_{i = 1}^{N}  F_{(i)}^2(t) e^{2 \sigma_{(i)}(u)}d l_{(i)}^2.
\end{equation}
In general the field equations are very cumbersome, except for the case where  $u$ is  harmonic (\ref{harmonic radial coordinate}) and the metric functions $\gamma(u)$, $\beta(u)$, $\xi(u)$, $\sigma_{(i)}(u)$ are the same as in the static solution (\ref{static solution}). In this case, from the field equations $R_{22} = R_{33} = \cdots = R_{DD} = 0$ $(D = 4 + N)$ we find
\begin{equation}
\label{the functions of time}
\frac{\dot{B}}{B} \propto \frac{\dot{C}}{C} \propto \frac{{\dot{F}}_{(i)}}{F_{(i)}} \propto \frac{1}{V}, \;\;\;V = A B C \prod_{i = i}^{N}F_{(i)}, 
\end{equation}
where dots denote differentiation with respect to $t$.
Thus, without loss of generality we can set $A = f^{ q_{1}}$, $B = f^{ q_{2}}$, $C = f^{ q_{3}}$, $F_{(i)} = f^{ q_{i}}$, where $f$ is some function of $t$ and the $q$'s are constants. Substituting these back into the field equations  $R_{22} = R_{33} = \cdots = R_{DD} = 0$ we get a second order differential equation for $f$ whose solution is
\begin{equation}
\label{equation for f}
f(t) = \left(k_{1} t + k_{2}\right)^{1/Q}, \;\;\;Q = \sum_{j = 1}^{D} q_{j},
\end{equation}
where $k_{1}$ and $k_{2}$ are constants of integration. Clearly, $k_{2}$ may be turned to zero by choosing the zero point of the $t$ coordinate. 
Now, if we denote $p_{i} = q_{i}/Q$, and $k _{1} = k$, then the solution takes the form

\begin{equation}
\label{separable solution}
dS^2 =   e^{2a u} dt^2 - (k t)^{2 p_{1}}\; e^{2 (a + b + c + s) u}\;  du^2 - (k t)^{2 p_{2}}\; e^{2 c u} dz^2 - (k t)^{2 p_{3}}\; e^{2 b u} d\phi^2                             - \sum_{i = 1}^{N} (k t)^{2 p_{(i + 3)}}\; e^{2 s_{(i)} u}d l_{(i)}^2,
\end{equation}
where the coefficients $p_{i}$ have to satisfy the Kasner-like relations
\begin{equation}
\label{Kasner-like behavior}
\sum_{i = 1}^{D} p_{i} = \sum_{i = 1}^{D} p_{i}^2 = 1.
\end{equation}
The parameters $a, b, c, s_{(i)}$ satisfy (\ref{compatibility condition}) as before; but in addition the cross component $R_{01} = 0$ now imposes an additional condition, viz., 
\begin{equation}
\label{relation between the p's and a, b, c, s}
(a - b - c - s) p_{1} + c p_{2} + b p_{3} + \sum_{i = 1}^{N} s_{(i)} p_{(i + 3)} = a.
\end{equation}
For any given set  $(a, b, c, s_{(i)})$ this is an additional condition on the Kasner parameters $p_{k}$. Consequently, the number of independent $p$'s is equal to $N$, the number of extra dimensions. In particular, in $4D$ these parameters are completely determined by (\ref{Kasner-like behavior}) and (\ref{relation between the p's and a, b, c, s}).

 Now, after the coordinate transformation (\ref{transformation of coordinates}), the line element (\ref{separable solution}) becomes
\begin{equation}
\label{Levi-Civita-Kasner  in 4 + N}
dS^2 =  {\hat{r}}^{2 m} d{\hat{t}}^2 - {\hat{r}}^{\left[2 (m + w)(m - 1) + w^2 + {\bar{w}}^2\right]}\left({\hat{t}}^{2 p_{1}} d{\hat{r}}^2 + {\hat{t}}^{2 p_{2}}d{\hat{z}}^2\right) - {\hat{{\cal{C}}}}^2  {\hat{t}}^{2 p_{3}}\;  {\hat{r}}^{2(1 - m - w)} d\phi^2 - \sum_{i = 1}^{N} {\hat{t}}^{2 p_{(i + 3)}}{\hat{r}}^{2 w_{(i)}} \; d {\hat{l}}_{(i)}^2,
\end{equation}
where the conicity parameter is now given by (for brevity of the presentation we omit the relationship between $(t, r, z, l_{(i)})$) and $(\hat{t}, \hat{r}, \hat{z}, {\hat{l}}_{(i)})$)
\begin{equation}
\label{definition of C-hat}
{\hat{{\cal{C}}}} = k ^{\omega}\times \left(a + b + s\right)^{(b - a p_{3})/[a(1 - p_{1}) + b + c + s]}, \;\;\;\omega = \frac{(a + b + c + s) p_{3} - b p_{1}}{(1 - p_{1}) a + b + c + s}.
\end{equation}
For $p_{1} = p_{3} = 0$, it reduces to ${\cal{C}}$ in the static case. In the above expression it is assumed that the denominator is not zero. In the case where  $(1 - p_{1}) a + b + c + s = 0$, which requires $p_{1} \neq 0$ (see footnote $1$),  we find $k^{p_{1}} = (a + b + s)$ and 

\begin{equation}
\label{definition of C-hat, special case}
{\hat{{\cal{C}}}} = (a + b + s)^{p_{3}/p_{1}}.
\end{equation}
It is interesting to note that in the nonstatic case the conicity parameter depends explicitly on the expansion rates along $r$ and $z$. 
In $4D$ $(w_{(i)} = 0)$, when  $p_{1} = p_{2}$ our solution (\ref{Levi-Civita-Kasner  in 4 + N}) reduces to the Levi-Civita-Kasner spacetimes discussed in \cite{Delice}. 

\medskip

{\bf Interpretation in $4D$:} We now proceed to study the effective four-dimensional world. First, to establish the effective metric in $4D$  we use a factorization of the metric which is a standard technique in Kaluza-Klein theory.  For this, we recall that  $R_{(D)}$,  the curvature scalar associated with the metric 

\begin{equation}
\label{general metric}
dS^2 = \gamma_{\mu\nu}(x)dx^{\mu}dx^{\nu} - \sum_{i = 1}^{N}H_{i}^2(x)dy_{i}^2,
\end{equation}
 can be expressed as
\begin{equation}
\label{relation between the curvature invariants}
\sqrt{|g_{(D)}|}\; R_{(D)} \propto \sqrt{|g_{(4)}^{\mbox{eff}}|}\; R_{(4)} + \mbox{other terms},
\end{equation}
where  $R_{(4)}$ is the four-dimensional  curvature scalar calculated from the effective $4D$ metric tensor\footnote{For the effective action in $4D$ to contain the exact Einstein Lagrangian, i.e. to deal with a constant effective gravitational constant, $g_{\mu \nu}^{\mbox{eff}}$ should be identified with the physical metric in ordinary $4D$ spacetime \cite{Davidson Owen}, \cite{Dolan}.}
\begin{equation}
\label{effective metric}
g_{\mu \nu}^{\mbox{eff}} = \gamma_{\mu\nu}\; \prod_{i = 1}^{m} H_{i}; 
\end{equation} 
$g_{(D)}$ and $g_{(4)}^{\mbox{eff}}$ denote the determinants of the $D$-dimensional metric (\ref{general metric}) and effective $4D$ metric (\ref{effective metric}), respectively, and the ``other terms" are proportional to the Lagrangian of an effective energy-momentum tensor (EMT) $T_{\mu\nu}$ in $4D$.

Thus, in the static case the effective gravity in $4D$ is determined by the line element (in what follows we omit the tilde over the coordinates)

\begin{equation}
\label{effective static metric in 4D, compact extra dimensions}
ds^2 = r^{(2m + w)} dt^2 - r^{[2(m + w)(m - 1) + w^2 + {\bar{w}}^2 + w]}\;\left(dr^2 + dz^2\right) - {\cal{C}}^2\; r^{[2(1 - m) - w]} \; d\phi^2. 
\end{equation}
The effective EMT is given by 
\begin{equation}
\label{Effective EMT for the static case}
8 \pi T_{0}^{0} = \frac{w^2 + 2 {\bar{w}}^2}{R^2 \; [2 m^2 + 2 m(w - 1)  + 2 + w^2 + {\bar{w}}^2 - w]^2}, \;\;\;T_{1}^{1} = - T_{0}^{0}, \;\;\;T_{2}^{2} = T_{3}^{3} = T_{0}^{0}.
\end{equation}
where $R$ is the proper (physical) radius of a shell with coordinate $r$,   
\begin{equation}
\label{physical radius for the 4D effective static metric}
R = \int_{0}^{r}{\sqrt{- g_{\tilde{r} \tilde{r}}} \; d\tilde{r}}.
\end{equation}
We note that the denominator in (\ref{Effective EMT for the static case}), except for $R = 0$, does not vanish for any real $m$ and $w$.        For ${\bar{w}}^2 = 0$, which implies $w_{(i)} = 0$ and $w = 0$, the EMT vanishes and (\ref{effective static metric in 4D, compact extra dimensions}) reduces to the Levi-Civita spacetime.

The relationship between the components of the EMT suggest that the source can be interpreted as a neutral massless scalar field  
\begin{equation}
\label{scalar field}
\Psi = \int{\sqrt{- 2 g_{rr}T_{0}^{0}}\; dr}.
\end{equation}
After integration we find
\begin{equation}
\label{scalar field}
\Psi(r) = \frac{1}{4}\sqrt{\frac{w^2 + 2 {\bar{w}}^2}{\pi}}\; \ln {r}.
\end{equation}
It is not difficult to verify that $\Psi$ satisfies the Klein-Gordon equation,
which 
is consistent with our interpretation. 

The coordinate transformation ${\cal{C}} \rho = \ln{r}$
renders the metric (\ref{effective static metric in 4D, compact extra dimensions})  into a  form where                                            $g_{\rho \rho} = - g_{tt}g_{zz} g_{\phi \phi}$. In terms of $\rho$ the solution of the Klein-Gordon equation is $\Psi = q \rho$, where $q$ is interpreted as the scalar charge \cite{BronnikovActaPhys}. Thus, from (\ref{scalar field}) we find the scalar charge to be 

\begin{equation}
\label{scalar charge}
q = \frac{{\cal{C}}}{4}\sqrt{\frac{w^2 + 2 {\bar{w}}^2}{\pi}}.
\end{equation}
Thus, (\ref{effective static metric in 4D, compact extra dimensions})-(\ref{scalar charge}) is a static, cylindrically symmetric solution of the coupled Einstein-massless scalar field equations.

$\bullet$ However, the factorization technique is not the only way to establish the effective 4-dimensional picture. Other alternatives  for dimensional reduction  are formulated in  induced-matter and braneworld theories, where the extra dimensions are not assumed to be compact. Another possible interpretation in $4D$ of the higher-dimensional metric is provided by Campbell's theorem \cite{Campbell}-\cite{Dahia}, which serves as a ladder to go between manifolds whose dimensionality differs by one. This theorem, which is valid in any number of dimensions, implies that every solution of the $4D$ Einstein equations with arbitrary energy-momentum tensor can be embedded, at least locally, in a solution of  the five-dimensional vacuum Einstein field equations. 

The metric of the vacuum solution in $5D$ $(w^2 = {\bar{w}}^2)$ is
\begin{equation}
\label{static metric in 5D}
dS^2 = r^{2 m} dt^2 - r^{2[(m + w)(m - 1) + w^2]}\left(dr^2 + dz^2\right) - {{\cal{C}}}\; r^{2(1 - m - w)} d\phi^2 - r^{2 w} dl^2
\end{equation}
In this interpretation the metric induced on every $4D$ hypersurface locally orthogonal to the extra dimension $l$ is given by 

\begin{equation}
\label{static metric in 4D}
ds^2 = r^{2 m} dt^2 - r^{2[(m + w)(m - 1) + w^2]}\left(dr^2 + dz^2\right) - {{\cal{C}}}\; r^{2(1 - m - w)} d\phi^2.
\end{equation}
The components effective EMT measured by an observer living in $4D$, and not aware of the extra dimensions, are
\begin{eqnarray}
\label{ETM for Campbell's interpretation}
8 \pi T_{0}^{0} &=&  \frac{- m w}{R^2 [m^2 + w^ 2 +(w - 1)(m - 1)]^2}, \nonumber \\
T_{1}^{1} &=& - T_{0}^{0} \; \frac{m^2 + (w - 1)(m + w - 1)}{m}, \nonumber \\
T_{2}^{2} &=&  T_{0}^{0} \; \frac{m^2 + (w - 1)(m + w)}{m}, \nonumber \\
T_{3}^{3} &=&  - T_{0}^{0} \; \frac{m + w - 1}{m}.
\end{eqnarray}
We note that 
\begin{equation}
\label{trace of the static EMT }
T_{0}^{0} + T_{1}^{1} + T_{2}^{2} + T_{3}^{3} = 0,
\end{equation}
which means that the effective matter is radiation-like. Therefore, in the context of braneworld models the metric (\ref{static metric in 5D}) is a static solution of the {\it empty} field equations on the brane. The effective EMT is the dark energy (or Weyl radiation) coming from the bulk to the brane, which is described by the projection of the $5D$ Weyl tensor on the brane.  

We note that both interpretations lead not only to different EMT, but also to different tangential velocities for test particles moving along  circular geodesics ($r = $ constant, $z = $ constant).  
The angular velocity $\omega$ of a particle moving along a circular geodesic is $\omega = \dot{\phi}/\dot{t}$, , where the dot stands for differentiation with respect to some affine parameter, and its tangential speed $V$ is given by $V = \omega/\sqrt{g_{tt}}$. After a simple calculation we find 
\begin{equation}
\label{tangential velocity}
V^2_{KK} = \frac{m + w/2}{1 - m - w/2}, \;\;\;V^2_{C} = \frac{m}{1 - m - w},\;\;\; w = \sum_{i}^{N} w_{(i)},
\end{equation}
where $V_{KK}$ and $V_{C}$ (here KK stands for Kaluza-Klein and C for Campbell) represent the tangential speeds calculated with the         metrics (\ref{effective static metric in 4D, compact extra dimensions}) and (\ref{static metric in 4D}), respectively\footnote{It is easy to prove that the expression for $V_{C}$ in (\ref{tangential velocity}) is in fact valid for any number of extra dimensions, not only in $5D$. }. It is important to note that when  $m \geq 0$, the physical constraint $V^2 \leq 1$  automatically assures the fulfillment of  $({1 - m - w/2}) > 0$ and $(1 - m - w) > 0$, which are required for the interpretation of $\phi$ as an angular coordinate\footnote{By analogy with flat spacetime, we assume that the angular metric coefficient  vanishes at  $r = 0$.}. 

From (\ref{tangential velocity}) it follows that $|V_{C}| >|V_{KK}|$, when $w < 0$ (and $m > 0)$, in the whole range $- 2m \leq w \leq 0$ where $V_{KK}^2 > 0 $. The opposite, i.e.  $|V_{C}| < |V_{KK}|$, occurs when $w > 0$ and $m > 0$. This is consistent with the fact  that  $T_{0}^{0}$ in (\ref{ETM for Campbell's interpretation}) is positive for $w < 0$ and negative\footnote{The effective matter quantities (\ref{ETM for Campbell's interpretation}) do not have to satisfy the regular energy conditions, or any physically motivated equation of state, because they involve terms of  geometric origin \cite{Bronnikov}.} for $w > 0$ $(m > 0)$, as well as with the Newtonian notion that for a test particle to remain in a circular orbit an increase (decrease) of the effective energy density, per unit length,   requires an increase (decrease)
of tangential speed  in such a way  that the centrifugal force can balance the gravitational attraction. 
If $w = 0$, then $|V_{C}| =  |V_{KK}| = \sqrt{m/(1 - m)}$ as in the Levi-Civita spacetimes (\ref{Levi-Civita spacetime}). To avoid misunderstanding, it should be emphasized that $w = 0$ ($w = \sum_{i}^{N} w_{(i)} = 0)$ does not imply that the extra dimensions are flat. 

The above  dimensional reduction techniques can be applied to the Levi-Civita-Kasner solutions (\ref{Levi-Civita-Kasner  in 4 + N}).  However, in view of the number of parameters the analysis is much more elaborated. Therefore, we defer the discussion of the effective   nonstatic  $4D$ spacetimes to another place.

\medskip

To finish this paper we would like to note that the solutions (\ref{Levi-Civita in 4 + N}) and (\ref{Levi-Civita-Kasner  in 4 + N}) can be easily extended to a chain of several Ricci-flat internal spaces. This requires everywhere the replacement

\begin{equation}
\label{definition of dg's}
dl_{(i)}^2 \rightarrow  \sum_{a,b = 1}^{n_{(i)}}\delta_{a b}(x_{_{(i)}})\;  dx_{(i)}^{a}dx_{(i)}^{b}, \;\;\;\;w = \sum_{i = 1}^{N} w_{(i)} \rightarrow \sum_{i = 1}^{N}n_{(i)} w_{(i)},\;\;\; {\bar{w}}^2 = \sum_{i = 1}^{N} w_{(i)}^2 \rightarrow \sum_{i = 1}^{N}n_{(i)} w_{(i)}^2
\end{equation}
where $n_{(i)}$ is the dimension of the $i$-th subspace with metric $\delta_{a b}(x_{(i)})$, and  the Ricci tensors $R_{a b}(x_{(i)})$ formed out by the $\delta_{a b}(x_{(i)})$ all vanish.

\end{document}